TRANSFORMING HIGH SCHOOL PHYSICS WITH MODELING AND COMPUTATION

by

JOHN M. AIKEN

Under the Direction of Brian D. Thoms

ABSTRACT

The Engage to Excel (PCAST) report, the National Research Council's Framework for K-12 Science Education, and the Next Generation Science Standards all call for transforming the physics classroom into an environment that teaches students real scientific practices. This work describes the early stages of one such attempt to transform a high school physics classroom. Specifically, a series of model-building and computational modeling exercises were piloted in a ninth grade Physics First classroom. Student use of computation was assessed using a proctored programming assignment, where the students produced and discussed a computational model of a baseball in motion via a high-level programming environment (VPython). Student views on computation and its link to mechanics was assessed with a written essay and a series of think-aloud interviews. This pilot study shows computation's ability for connecting scientific practice to the high school science classroom.

INDEX WORDS: Computational Modeling, Physics, Education

TRANSFORMING HIGH SCHOOL PHYSICS WITH MODELING AND COMPUTATION

by

JOHN M. AIKEN

A Thesis Submitted in Partial Fulfillment of the Requirements for the Degree of

Masters of Science

in the College of Arts and Sciences

Georgia State University

2013

TRANSFORMING HIGH SCHOOL PHYSICS WITH MODELING AND COMPUTATION

by

JOHN M. AIKEN

Committee Chair:     Brian D. Thoms

Committee:    Marcos D. Caballero

Michael F. Schatz

Raj Sunderraman

Electronic Version Approved:

Office of Graduate Studies

College of Arts and Sciences

Georgia State University

December 2013

## ACKNOWLEDGEMENTS

I would like to acknowledge everyone who has participated in providing me with a transformative experience in my Masters program. Mike Schatz and Brian Thoms have fulfilled complimentary roles as advisers. Danny Caballero and Shih-Yin Lin have given me valuable research advice. I'd also like to acknowledge Scott Douglas, whose aid in coding and quantifying data has been enlightening, and Chastity Aiken, my wife, whose own expertise in research has been as valuable as her personal advice and support.

# TABLE OF CONTENTS

## LIST OF TABLES

# LIST OF FIGURES

# 1 INTRODUCTION AND BACKGROUND

## 1.1 Transforming High School Physics for the 21st Century

The high school physics classroom of the 21st century is failing at preparing students for undergraduate studies and the work place [1-3]. This is because, in part, the average classroom does not focus on teaching scientific practices (like computation, modeling, communicating, etc.) and instead stresses content acquisition and "fact learning" [3]. When surveyed, over 400 employers report that high school students possess none of the skills they expect new employees to possess. Employers rated four skills as most important: work ethic, oral and written communications, the ability to collaborate, and critical thinking (see Table 1 for the full list); when surveyed, high school graduates are inadequate in all of these skills [2]. Further, the Engage to Excel report to the President of the United States (PCAST report) estimates that there will be approximately three million new Science, Technology, Engineering, and Mathematics (STEM) jobs opening in the next decade [1]. The PCAST report also projects that the current production of STEM professionals will only fill about two thirds of the available jobs. This failure to prepare an adequate number of STEM majors for STEM jobs is, in part, because only ~40% of declared STEM majors finish their STEM undergraduate programs [1]. When surveyed, high-achieving students most often cite "uninspiring introductory courses" as their reason for dropping out [1]. The PCAST report recommends five ways to keep and attract people to STEM degree programs:

*(1) Catalyze widespread adoption of empirically validated teaching practices;*

*(2) Advocate and provide support for replacing standard laboratory courses with discovery based research courses;*

*(3) Launch a national experiment in postsecondary mathematics education to address the mathematics preparation gap;*

*(4) Encourage partnerships among stakeholders to diversify pathways to STEM careers; and*

*(5) Create a Presidential Council on STEM Education with leadership from the academic and business communities to provide strategic leadership for transformative and sustainable change in STEM undergraduate education.*

Of the five paths to achieving the PCAST report's goal, high school physics courses are already well equipped to tackle the first two. The Physics Education Research community has a long history of informing the pedagogy and curricula of high school physics classrooms. These instructional innovations include research-validated curriculum design, lessons learned from cognitive science, new tools for students to use, and systematic measurements of affect, attitude, and efficacy [4-10]. But what does the anatomy of science education reform look like? And how will it be specifically delivered to science teachers?

The National Research Council's Framework for K-12 Science Education declares that the most effective way to teach science is to have students learning the practices of a professional scientist [3]. The framework highlights three major dimensions in which science education should be better aligned with science practice:

*(1) Scientific and engineering practices*

*(2) Crosscutting concepts that unify the study of science and engineering through their common application across fields*

*(3) Core ideas in four disciplinary areas: physical sciences; life sciences; earth and space sciences; and engineering, technology, and applications of science*

Scientific practices include critical thinking by "asking questions and defining problems", building communication skills by "constructing explanations and designing solutions", and using computation and model building to solve real world problems [3]. The framework's emphasis on making science class more like professional science practice motivated much of the research and classroom implementation found in this thesis. This work focuses chiefly on the introduction of two scientific practices into the high school physics classroom: computation and model building.

**Table 1 : Skills sought after by top employers [2].**

| Basic Knowledge/Skills | Applied Skills |
|---|---|
| English Language (spoken) | Critical Thinking/Problem Solving |
| Reading Comprehension (in English) | Oral Communications |
| Writing in English (grammar, spelling, etc.) | Written Communications |
| Mathematics | Teamwork/Collaboration |
| Science | Diversity |
| Government/Economics | Information Technology Application |
| Humanities/Arts | Leadership |
| Foreign Languages | Creativity/Innovation |
| History/Geography | Lifelong Learning/Self Direction |
| | Professionalism/Work Ethic |
| | Ethics/Social Responsibility |

How, then, do professional scientists practice science? That is, can real scientific practice (specifically computation and model building) be characterized in a consistent way? And how do these practices compare to what students do in an introductory physics course? In his seminal work with the Maryland University Project in Physics and Educational Technology (M.U.P.P.E.T.), Edward Redish reported that introductory college physics courses "introduce[d] students to the basic content of physics, [but they] provide[d] almost no activities that illustrate how research is done," [4]. Redish's comparison between the practices of students and the practices of a professional scientist show a stark reality; science class does not resemble professional science (see Table 2). In class, students are given laws and rules from on high, and have personal ownership of them; scientists tweak and play with models in their attempts to describe physical phenomena, demonstrating familiarity with and ownership of the models.

Students attempt to solve analytically tractable problems; scientists most often address problems which have no analytic solution, and so they use numerical methods. Thus, Redish claims students are not using tools like computers to solve meaningful problems in the same way that scientists do.

**Table 2: Comparison of students' activities with those of a professional physicist [4].**

| Students: | Professional Scientists: |
|---|---|
| Solve narrow, pre-defined problems of no personal interest | Solve broad, open ended and often self discovered problems |
| Work with laws presented by experts. Do not "discover" them on their own or learn why we believe them. Do not see them as hypotheses for testing. | Work with models to be tested and modified. Know that "laws" are constructs. |
| Use analytic tools to get "exact" answers to inexact models. | Use analytic and numerical tools to get approximate answers to inexact models. |
| Rarely use a computer. | Uses computers often. |

Students also know that the science classroom does not resemble professional science. When surveyed about physics' "real-world connections" using the Colorado Learning Attitudes about Science Survey (CLASS), students demonstrate a significant difference in their views and the views of professional physicists [5]. The CLASS aims to measure "student's beliefs about physics and learning physics" [5]. Consisting of 36 items concerning topics likes problem solving (confidence, sophistication, etc.), real-world connections, and personal interest, the CLASS has shown that overall, physics instruction has a *negative impact on student's attitudes towards physics*. Furthermore, physics instruction has a more significant impact on women's attitudes towards physics. Some instructional reformers have emphasized to need to highlight "real-world connections" to create positive attitudes and increased conceptual understanding, but what does this look like in the classroom [1, 3]?

The Investigative Science Learning Environment (ISLE) labs exposed students in an introductory physics course to an environment that encouraged scientific thinking and practices [6]. ISLE provides instructors with course structure guidelines, lab activities, and rubrics. ISLE focuses on helping students develop scientific reasoning. Students in this course respond positively. Students participating in an introductory college physics lab designed to teach practices valuable in the workplace exhibited expert-like sense making and experimental design behavior [6]. These results could not be replicated in a tradi-tional laboratory, even when the laboratory curriculum was augmented with conceptual and reflection questions.

## 1.2    Building on a Research Base: Modeling and Computation

Changing how physics is taught,such as ISLE, can lead students to become more scientist-like in their thinking. Such changes give students the opportunity to practice using new tools to solve problems (like computation and modeling), and expand their ability to communicate and work with others. This work attempts to make high school science more like real science in two ways:

1. Students will build predictive models of physical phenomena that they can observe in their world.

2. Students will actively engage in building computational simulations and modeling to solve these problems.

This work focuses on the NRC framework's Science and Engineering practices dimension [3]. While there are eight total scientific practices emphasized by the NRC, two of these practices stand out: developing and using models, and using computational thinking (see Table 3). It is important to align our goals with the NRC framework for two reasons: (1) Education transformation should be focused on re-research-validated practices [1], (2) Next Generation Science Standards (NGSS) are based on the NRC framework [7].

Table 3: NRC Framework for K-12 Science Education [3].

| The Three Dimensions of the Framework | |
|---|---|
| Scientific and Engineering Practices | Asking questions (for science) and defining problems (for engineering) |
| | Developing and using models |
| | Planning and carrying out investigations |
| | Analyzing and interpreting data |
| | Using mathematics and computational thinking |
| | Constructing explanations (for science) and designing solu- |

| | | |
|---|---|---|
| | | tions (for engineering) |
| | | Engaging in argument from evidence |
| | | Obtaining, evaluating, and communicating information |
| Crosscutting Concepts | | Patterns |
| | | Cause and effect: Mechanism and explanation |
| | | Scale, proportion, and quantity |
| | | Systems and system models |
| | | Energy and matter: Flows, cycles, and conservation |
| | | Structure and function |
| | | Stability and change |
| Disciplinary Core Ideas | | *Physical Sciences* |
| | | PS1: Matter and its interactions |
| | | PS2: Motion and stability: Forces and interactions |
| | | PS3: Energy |
| | | PS4: Waves and their applications in technologies for information transfer |
| | | |
| | | *Life Sciences* |
| | | LS1: From molecules to organisms: Structures and processes |
| | | LS2: Ecosystems: Interactions, energy, and dynamics |
| | | LS3: Heredity: Inheritance and variation of traits |
| | | LS4: Biological evolution: Unity and diversity |
| | | Earth and Space Sciences |
| | | ESS1: Earth's place in the universe |
| | | ESS2: Earth's systems |
| | | ESS3: Earth and human activity |
| | | |
| | | *Engineering, Technology, and Applications of Science* |
| | | ETS1: Engineering design |
| | | ETS2: Links among engineering, technology, science, and society |

Computation and modeling were singled out from among all the NRC practices because they feature prominently in the Modeling Instruction (MI) curriculum. The MI curriculum has a long history of research supporting its success at core content acquisition through teaching procedural modeling building to students [8-10],. The MI curriculum emphasizes model building, experimentation, and using scientific practices in the classroom. MI will be described thoroughly in Section 1.3. Computation has a long history of education research both in and out of physics[11-14]. In physics education research specifically, computation is often taught with the Python programming language augmented with the Visual module (the language/module pair is colloquially called "VPython"). VPython gives students access to animations, graphs, and a host of other scientific tools [15]. In addition, another Python module named PhysUtil provides specific tools that link VPython to the MI curriculum [16], and is used extensively in computational work described here. VPython will be described in Section 1.4.

**1.3      Modeling Instruction for High School Physics**

Traditional physics education consists of lectures, disconnected labs, and little interaction between students and their peers or with faculty [17]. The Modeling Instruction (MI) curriculum was created to change high school physics class into a place where students explored physical phenomena with experimentation and systematic analysis [8]. MI aims to help students achieve four specific goals [18]. These goals are:

1.   make sense of a physical experience,

2.   understand scientific claims,

3.   articulate coherent opinions of their own and defend them with cogent arguments, and

4.   evaluate evidence in support of justified belief.

To facilitate these goals in the classroom, students are tasked with testing "scientific models" that describe physical phenomena. The modeling process is foreign to the traditional classroom setting [8]. Therefore, teachers were trained in the MI curriculum in workshops. Activities were created to explore each model described within the MI curriculum; these activities are available on the Modeling Instruction website free of charge [19].

**1.3.1      *What Is a Scientific Model?***

There is no unique definition of the word "model" in the scientific literature [10]. The authors of MI agree that a "model" has a "well defined scope" [10]. When queried, scientists define models as general frameworks that govern specific behavior found in nature, but not as rigorously as a theory [10].

The MI authors define a scientific model as a "conceptual system mapped, within the context of a specific theory, onto a specific pattern in the structure and/or behavior of a set of physical systems so as to reliably represent the pattern in question and serve specific functions in its regard" [10]. As a practical example, the "conceptual system" might be a physical concept like "constant velocity", where the behavior is dictated by a stationary or moving object that does not exhibit any acceleration. This model can be represented with equations (e.g., $\vec{v} = \dfrac{\Delta \vec{r}}{\Delta t}$), graphs, and diagrams; a plurality of representations is a general feature of models. A constant velocity model serves a very specific function; it cannot explain, for example, the motion of an object experiencing a non-zero net force, so another model is required to explain this behavior.

Scientists build and use models in a similar way, with similar limitations. The Hodgkin-Huxley model for action potentials in neurons describes the biophysical behavior of cell membranes. This model was a very good model, earning Hodgkin and Huxley the Nobel Prize in "Physiology or Medicine" in 1963 [20], but it made assumptions that ended up failing under experimental investigation and thus has been updated and expanded [21]. Likewise, plate tectonics, the model that describes how the Earth's surface moves, originally did not include a motivation for this movement. It wasn't until the 1960s that mantle convection provided evidence for what motivated plate movement [22]. In both cases, scientists started with a model, attempted to verify its explanation via experiment or observation, and changed the model or created a new one when the current model did not accurately explain the observed phenomena. The modeling process in MI is built to align with professional research. This process is called the "modeling cycle" [9].

### 1.3.2 Modeling Cycle

The "modeling cycle" is broken into two major steps: (1) the "Model Development" stage, and (2) the "Model Deployment" stage. Students begin the Model Development stage by learning about a new model via a pre-lab demonstration and discussion. They then participate in a "paradigm experiment", exploring a particular physical phenomenon with experiment and observation in small groups. This is followed with a post-lab discussion. Students present their results to the class and participate in a discussion of the failings or successes of each group's experiment. The experiment/observation process of the Model Development stage is followed by the Model Deployment stage. In this stage, the students are tasked with completing worksheets that ask students to further develop concepts by interacting with various representations of the model (e.g., graphs, equations, diagrams (see Figure 1), etc.). These worksheets are usually completed in small groups [9]. Students also are given quizzes that ask conceptual and quantitative questions. The final assessment in the Model Deployment stage is a lab practicum and a test. The lab practicum asks the students to design an experiment that uses the model they have been studying to solve a real world problem. This real-world problem often asks students to extend their knowledge. For instance, a common lab practicum for the constant velocity model has students predicting the position where two constant-velocity carts collide, assuming they travel at different speeds [19, 23]. This practicum has students practicing data gathering, using their models to predict future behavior, and exposes them to the idea of momentum transfer, which cannot be explained with a model describing only constant velocity. Before being introduced to a new model, students are given a summative assessment exam that assesses their problem solving. MI is supported by free materials hosted on the American Modeling Teachers Association (AMTA) website, as well summer modeling workshops [8,18,19]. These materials are comprehensive, and include learning goals, activities, and experiments. An example of these materials is detailed below.

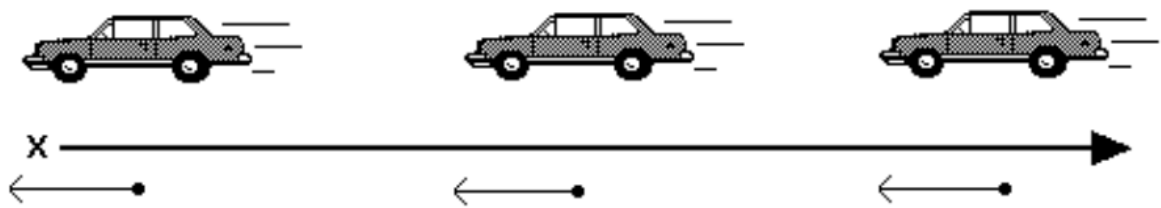

Figure 1: A motion diagram of a car moving at constant velocity. This image is used in a Modeling Instruction worksheet [24].

### 1.3.3    An Example of a Modeling Instruction Model: Constant Velocity

The recently updated (2010) constant velocity unit introduces students to the idea of building models, and to their experimental tools [23]. The unit begins with an explanation of the instructional goals of the unit aimed at the instructor. These goals concern students' use of coordinate systems, the mathematical vernacular of kinematics, and an introduction to the software used for experimental design. A full description of these goals can be found in Table 4.

Table 4: Instructional goals of the constant velocity model unit. The particle model describes particle motion, i.e. it assumes a moving object is a point mass.  The original activity used inconsistent vector notation, where velocity was represented as a vector but position was not.  This is corrected here.

| Instructional Goals | Description |
|---|---|
| Reference frame, position and trajectory | Choose origin and positive direction for a system<br> Define motion relative to frame of reference<br> Distinguish between vectorial and scalar concepts<br> (displacement vs distance, velocity vs speed) |
| Particle Model | Kinematical properties (position and velocity) and laws of motion<br> Derive the following relationships from position vs time graphs:<br>$\Delta \vec{x} = \vec{x}_\mathrm{f} - \vec{x}_0$<br>$\vec{v} = \dfrac{\Delta \vec{x}}{\Delta t}$<br>$\vec{x} = \vec{v}t + \vec{x}_0$<br>$\Delta \vec{x} = \vec{v}t$ |
| Multiple representations of behavior | Introduce use of motion map and vectors<br> Relate graphical, algebraic and diagrammatic representations. |
| Dimensions and units | Use appropriate units for kinematical properties<br> Dimensional analysis |

Students begin the Model Discovery phase by exploring the motion of a battery-powered constant velocity car (CV car). They chart the motion of this activity on graph paper. They then determine the average velocity and displacement of the CV car in two ways: they find the slope of the graph they created first analytically (via rise over run, etc.), then algebraically (with the equation $v = \dfrac{x_f - x_0}{t_f - t_0}$). This is followed by classroom discussion about the observations they made and how they described these observations with graphs and mathematics. Next, the students attempt to produce experimental data that resembles graphs given by the instructor on worksheets (see Figure 2).

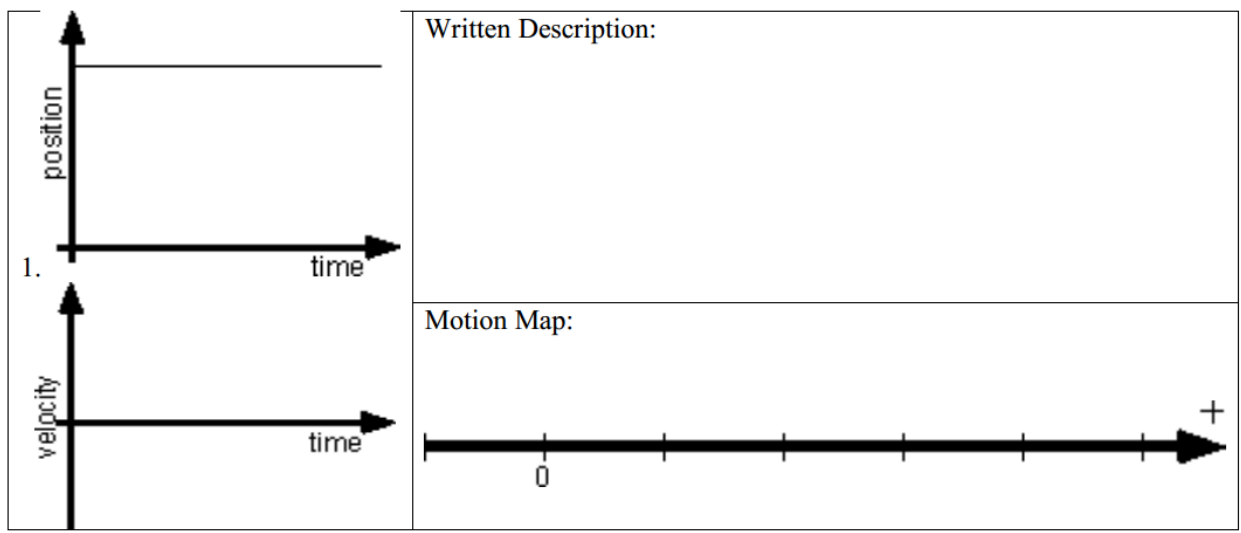

**Figure 2: Students are asked to move in front of a motion detector to reproduce the position versus time and motion map represented in this image.**

By the end of the Model Development phase, students have completed a paradigm lab, completed a reading assignment, and participated in follow-up exercises (see Table 5). They then begin Model Deployment. Model deployment begins with a quiz asking the students to make calculations based on motion maps. Motion maps are diagrams that represent motion with vectors, sometimes accompanied by a picture (see Figure 1). They then participate in the constant velocity lab practicum. This lab practicum has the students observing the motion of two constant velocity cars: one slow, one fast. They are tasked with predicting the position where the two cars will meet if they are driven towards each other. This lab activity is followed by more worksheets and quizzes culminating in an exam that covers all of the material.

Table 5 : Constant velocity model sequence.

| Model Stage | Description |
| --- | --- |
| Model Development | Constant velocity car motion lab |
| | Reading on motion maps |
| | Lab: Multiple representations of motion |
| | Worksheets: motion maps, position vs. time graphs, velocity vs. time graphs |
| Model Deployment | Quiz 1: Quantitative motion maps |
| | Constant velocity lab practicum: Dueling Cars |
| | Worksheet: position vs. time graphs, velocity vs. time graphs |
| | Quiz 2: average speed |
| | Worksheets: velocity vs. time graphs and displacement, multiple representations of motion |
| | Review sheet |
| | Constant velocity Test |

At the end of each modeling cycle, students are introduced to a new phenomenon that cannot be explained with their current model. Scientists go through a similar experience when their models fail to predict new phenomena. Not only does this process explicitly attempt to mirror the actions that professional scientists take to produce a model, it has produced noticeable differences in student outcomes. This will be discussed in the following section.

### 1.3.4    *Effectiveness of Modeling Instruction*

Modeling Instruction has had a strong relationship with Physics Education Research [9, 10, 25-30]. When compared to traditional lecture courses, Modeling Instruction has produced substantial content gains as measured by the Force Concept Inventory [8, 9, 26]. Students participating in Modeling courses taught by novice teachers (defined as teachers teaching MI for their first year) who predominately did not come from a physics background (i.e., their undergraduate education was in something other than physics) had scores 10% higher on post test (FCI) when compared to students in a traditional lecture physics classroom (see Figure 3) [9]., Eleven expert teachers were identified after their second year of teaching modeling. They were defined as experts not in terms of their physics knowledge, but in terms of how well they understood the modeling process. These teachers showed a 69% post test score average (see Figure 3). MI has also shown a positive impact on student affect and attitude. Introductory college physics students participating in a course structured around the MI curriculum were queried about their attitudes towards learning and doing physics using the Colorado Learning Attitudes about Science Survey (CLASS) [5, 25]. Researchers saw large positive shifts in student attitude towards learning and doing physics, particularly in the categories of problem-solving confidence, problem-solving sophistication, and applied conceptual understanding (note: these are attitudes towards decisions made to solve problems, not a measurement ability to solve problems) [25]. Students in traditional courses generally lose self-efficacy (a person's belief of their ability to complete a task) in physics [28, 31, 32]. In MI courses, students' scores change very little on pre/post self efficacy measures. While this initially seems to be a null result, no change in overall self-efficacy is a positive outcome for students compared to the negative change in traditional courses. Physics First classrooms also show positive results when using MI over traditional instructional methods. The name "Physics First" expresses the notion that the traditional high school order of science (biology, chemistry, physics) is backward, because physics is the foundational science upon which chemistry and biology are built [33]. Students in MI based 9th grade classes scored comparatively to their 12 grade students on the FCI, and considerably higher than non-MI peers [33]. Moreover, honors-track ninth graders in modeling classes scored significantly higher on post-FCI scores when compared to honors-track eleventh and twelfth grade non-modeling students (see Table 6) [30].

**Table 6 : Results from two Physics First and Modeling Instruction studies. L. Liang's study had a total of 301 participating students in grades 9-12 [30]. M. O'Brien's study had a total of 321 students participate [33]. M. O'Brien did not report standard deviation and reported average scores not percentages (represented in parantheses).**

| Class | Pre-Test Mean and SD | Post-Test Mean and SD |
| --- | --- | --- |
| *Modeling in Physics First (PF)* | | |
| **L. Liang, et. al.** | 24.80±7.37 | 47.50±6.57 |
| **M. O'Brien, et. al.** | (5.0) 18.52 | (8.9) 46.30 |
| **M. O'Brien, et. al. (honors)** | (4.9) 18.15 | (12.5) 46.30 |
| *Modeling in non-PF* | | |
| **L. Liang, et. al.** | 25.18±9.46 | 47.33±7.65 |
| *Non-modeling in PF* | | |
| **M. O'Brien, et. al.** | (5.6) 20.37 | (6.3) 48.15 |
| **M. O'Brien, et. al. (honors)** | (5.5) 20.37 | (13.0) 48.15 |
| *Non-modeling in non-PF* | | |
| **L. Liang, et. al.** | 26.28±8.02 | 38.47±4.36 |
| **L. Liang, et. al. (honors)** | 28.69±9.18 | 41.95±6.57 |
| **O'Brien et. al.** | (6.0) 22.22 | (10.9) 40.37 |

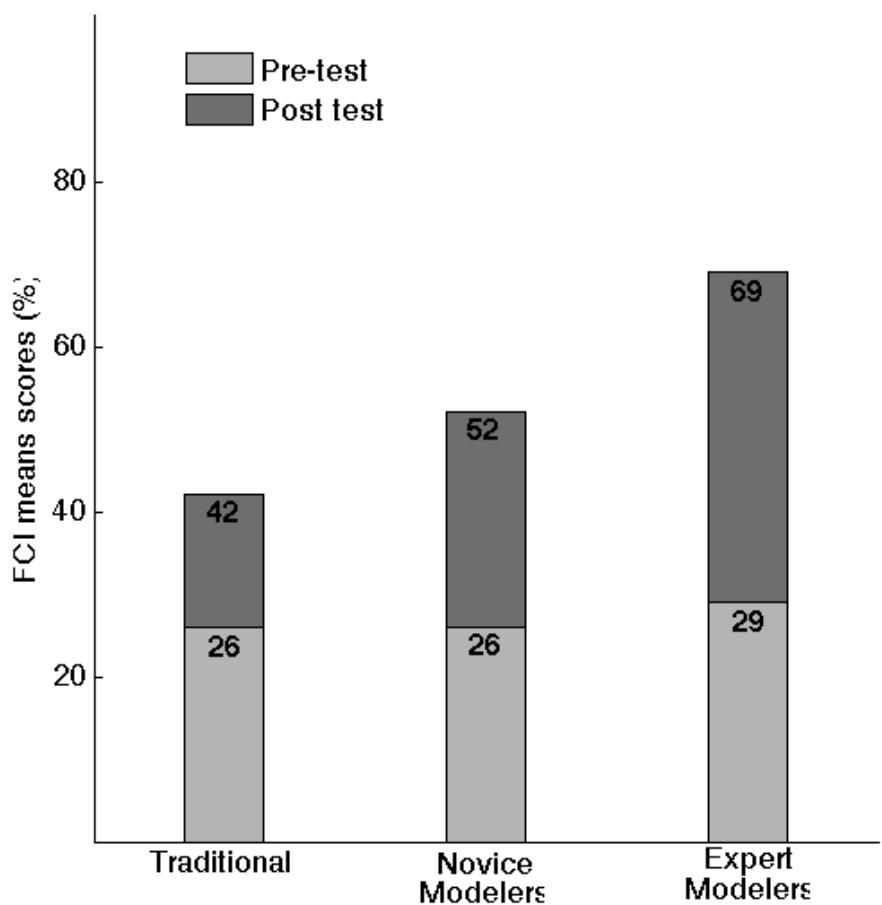

**Figure 3 : Student FCI mean scores. Modified from [9]. Jackson did not report error or significance in her study.**

### 1.3.5 Modeling Instruction's Goals for Teachers

Modeling Instruction's long history of research makes it an ideal candidate to meet the PCAST report's first two goals [1]. MI is an example of "empirically validated teaching practices" (see section 1.3.4). It also has been designed specifically for students to learn by discovery and doing. In fact, the American Modeling Teachers Association lists seven goals for teachers that align with both the PCAST report's goals for the classroom as well as the NRC framework's goals [18]. These teacher-aimed goals are:

(1)     Ground [teachers'] teaching in a well-defined pedagogical framework [here, Modeling Theory], rather than following rules of thumb;

(2)     Organize course content around scientific models as coherent units of structured knowledge;

(3)     Engage students collaboratively in making and using models to describe, to explain, to predict, to design and control physical phenomena;

(4)     Involve students in using computers as scientific tools for collecting, organizing, analyzing, visualizing, and modeling real data;

(5)     Assess student understanding in more meaningful ways and experiment with more authentic means of assessment;

(6)     Continuously improve and update instruction with new software, curriculum materials and insights from educational research;

(7)     Work collaboratively in action research teams to mutually improve their teaching practice.

Over twenty years of development of the MI curriculum has yielded great success in the pursuit of these goals [9]. Modeling Instruction provides a framework for the type of transformation in the science classroom that the authors of the PCAST report, the NRC framework, and the report on employer expectations are seeking [1-3]. But what is missing from the "traditional" modeling curriculum? How can Modeling Instruction be augmented to follow the practice of professional science even more closely? Scientific practice has overwhelmingly begun to include computation. How can computation fit into the Modeling Instruction framework?

## 1.4    Prior Use of Computation in 20th Century Science Classes

In recent decades, computation has risen as a third tier of science practice equal in prominence with theory and experiment/observation. Computation has been used to model a vast variety of physical situations like Earth's climate, nuclear interactions, and aerodynamic systems [34-36]. The advent of the personal computer has only made this complex tool ubiquitous and accessible to the common person. The 21st century student is a "digital native" [37]. Students use computers for online homework systems, email, instant messaging, and a whole host of other activities. yet using computers to solve complex problems is largely absent in undergraduate physics curricula (and there is little evidence to show it being used at lower levels, e.g. high school physics classrooms) [11]. Introducing computational modeling into the undergraduate physic major curriculum has been slow going [11]. Students leaving undergraduate programs have essentially been teaching programming to themselves once they enter the workforce [38]. Most science courses that do incorporate computers expose students to—at most— spreadsheet calculators, "black-box" analysis tools, and word processors [1]. By introducing computational problem solving into the K-12 classroom, e.g. using a computational model to predict projectile motion, students are better prepared to enter science and engineering undergraduate programs that require rigorous understanding of programming, mathematics, and modeling [39].

At the introductory college physics level, the Matter & Interactions (M&I—not to be confused with Modeling Instruction, MI) curriculum has gained some traction at several prominent universities (Carnegie Mellon University, Georgia Institute of Technology, North Carolina State University, and Purdue University), [40]. The M&I curriculum uses a model-based approach reforming introductory me-chanics in a modern physics context. It is intended to be taught in a computational environment, with many problems in the text being unsolvable by analytic means alone. While the work done by the M&I curriculum is a substantial move to align classroom physics with professional physics practice, only a minority of universities has yet adopted it. At the upper division, there have been small attempts to integrate computation into the general mechanics and electricity and magnetism curriculum [11, 41, 42]. However this work, while being based in research pedagogy has not been studied extensively itself.

Research in computation's role in the K-12 classroom has been ongoing since the advent of LOGO in the 1960s. LOGO is a computer language developed in the 1960s to teach children programming skills. It's hallmark feature is the "turtle" that students can command to move, draw, and perform other functions. Decades of research in LOGO has shown that students who are exposed to computation become more creative in their original thoughts, generate higher achievement in meta-cognitive processing, enhanced effectance motivation (i.e., the desire to master one's environment), and higher motivation [13, 43-45]. Klahr and Carver found that given a well structured environment, cognitive transfer of problem solving skills could be facilitated using Logo [12]. LOGO's use in computational science has been negligible outside of the computational science of agent-based modeling [46].

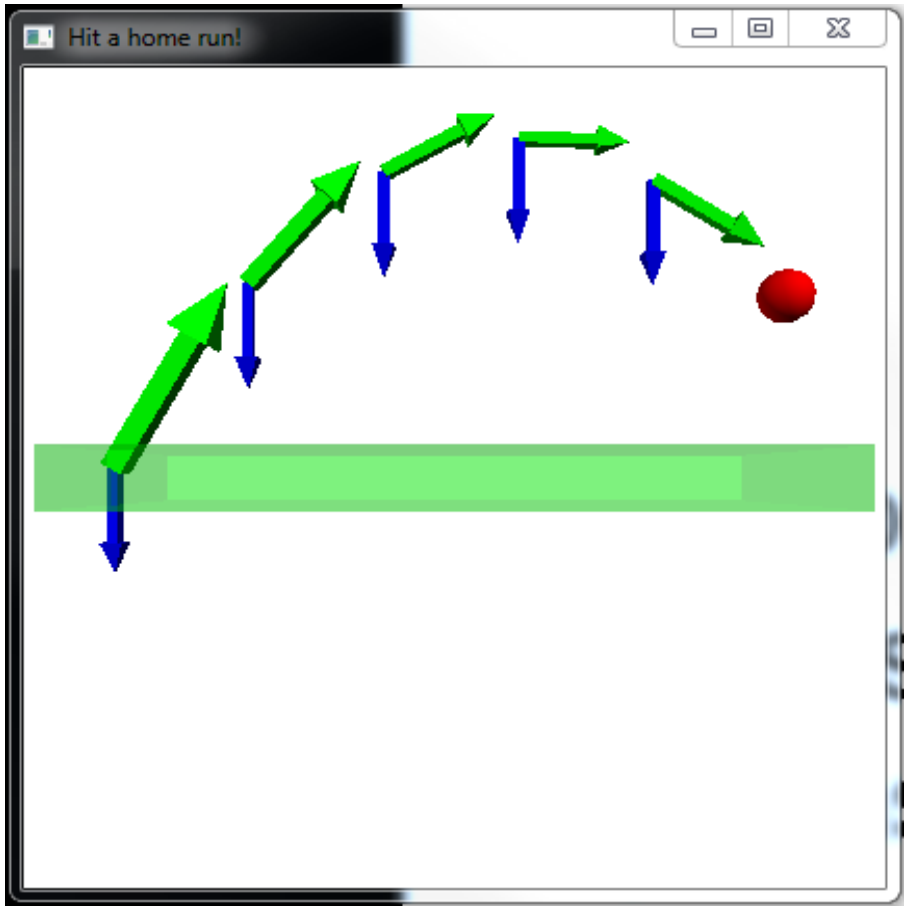

**Figure 4 : Example VPython screenshot of the animation of a baseball in flight. This animation would appear as a real time update of the integration calculation on a computer. Note the motion map and force vector arrows. These are sized by the magnitude of the vector they represent.**

Unlike LOGO, the Python programming language is used often in science for statistical and numerical analysis [47, 48]. It has also consistently been ranked in the top 10 most popular programming languages [49]. Python has been singled out by computer science education researchers as an ideal learning language given its clean syntax, dynamic typing, and the immediate feedback provided by the interpreter (see appendix 3.1 for a full list of reasons why Python is a good learning language). Modules have been specifically developed for Python to support physics education. The Visual module (colloquially known as VPython when paired with Python itself) has been in development for over a decade [14, 50]. VPython provides an easy syntax for creating dynamic animation and visualization of representa-tions that would otherwise be static in a traditional physics curriculum (see Figure 4). The visualization provided by a numerical model is of paramount importance; certain aspects of visualization help students communicate a more coherent picture of their understanding [51]. These graphical and diagrammatic descriptions of the physical phenomenon, which might otherwise form the sole basis of the students' exposure to the model, are reproduced precisely by the computational model. Furthermore, the linking of representations can be done quite easily with a few simple lines of code. VPython has been further extended by the PhysUtil module. The PhysUtil module has been specifically designed to provide students in a Modeling Instruction environment access to common model representations (graphs, motion maps, axes, etc.) [16]. Thus, VPython is the obvious choice of language for a high school introductory physics course that includes a computational modeling component.

## 1.5  Incorporating Computation into High School Physics

Incorporating computation into any curriculum can be a monumental task [11]. However, the motivation for doing so has been clearly provided by the NRC framework's report's call to include computation in the K-12 science education environment [3]. The PCAST report recommendation to utilize empirically validated teaching practices in conjunction with discovery-based research courses makes Modeling Instruction an obvious choice as a curriculum [1]. This is further reinforced by the NRC framework's recommendation to place model building and testing at the forefront of any science curriculum reform [3]. VPython's close relationship with physics education research (and its suite of tools built specifically to implement computation into Modeling Instruction—i.e., PhysUtil) distinguishes the language as the appropriate choice for the high school environment [15, 16]. But one piece is still missing: a pilot classroom.

Partnering with an in-service high school physics teacher at a K-12 co-educational independent day school located in Georgia), we began to develop computational activities for a 9th-grade Modeling Instruction course. To support the computational activities, each student had access to a laptop with VPython and PhysUtil installed. Leveraging the two-stage design of MI (see section 1.3.2), the instructor replaced some of the worksheet activities that would take place during the "Model Deployment" stage with computational assignments. Computational assignments would follow in-class experiments and problem-solving sessions. Beginning with the typical constant-velocity car experiment (see section 1.3.3), students then constructed a computational model of a constant-velocity car. Students used these computational models to make predictions for a variety of physical situations to which the model applied. Computation offers students the ability to easily solve every kinematics problem. This is due to requiring minute changes in initial conditions to represent a different problem.

Modeling Instruction treats each force and motion model as distinct, but the common thread of predicting motion using Newton's 2nd law and kinematics unites them. The computational algorithm used to predict motion likewise retains the distinctions between the force and motion models, but highlights the commonality among them: namely, that such models differ only in the net force exerted on the system and in their particular initial conditions. Given knowledge of the system's initial position and velocity, as well as the net force on the system, the algorithm for predicting motion can be described as a set of rules applied locally in space and time:

(1) At a given instant in time $t$, compute the net force, $\vec{F}_{net}$, acting on the system,

(2) For a short time $t$ later, compute the new velocity of the system using Newton's 2nd law,

(3) At the same new time ($t + \Delta t$), compute the new position of the object using this updated velocity, and

(4) Repeat Steps (1)-(3) starting at the updated time $t + \Delta t$.

Formally, the iterative application of Steps (1)-(3) is, in effect, explicit (Euler-Cromer) numerical integration of the equations of motion for Newtonian mechanics ($\Delta \vec{v} = \vec{a} = \dfrac{\vec{F}_{net}}{m}, \Delta \vec{r} = \vec{v} \Delta t$) [52]. Computation can be used thusly to highlight the instructional goals in the traditional MI curriculum for motion predic-tion (see section 1.3.3). In the constant velocity model, students have greater opportunity to see the links between the mathematical representation of kinematics and graphical representation when using computation. This is because the computational models can produce immediate graphical feedback.

The pedagogical advantages of computation are not limited to the constant velocity model. Computation highlights the relationship between the different physics models in MI (e.g., the no-forces model, the balanced-forces model, and the unbalanced-forces model). To produce simulations with qualitatively different behavior, students simply change the initial conditions (e.g., from 1D to 2D motion) or the net force (i.e., from constant to constantly changing). For example, the constant velocity/balanced forces model can be generalized to the constant acceleration/unbalanced forces model by inserting a constant net force into the computational model—this involves only a very straightforward change in the computer code. Furthermore, we can extend the constant acceleration/unbalanced forces model to parabolic motion model by giving the object an initial velocity in both the x and y directions.

Table 7 : Students were asked the question "Do you know anything about programming a computer?". If they responded yes they were asked to explain their experiences (N=36). One student who answered "yes" to the initial question responded in the follow up question explaining their experience with " Like downloading software from a CD". This was marked as "None" for answer purposes

| Experience | Number of students reporting |
|---|---|
| Only Q-BASIC in 7th grade | 14 |
| Summer experiences in C++ or JAVA | 3 |
| Arduino programming | 1 |
| None | 17 |
| No Answer to follow up question | 1 |

Students were given an informal technology experience survey at the beginning of the year. This survey had been given in previous years to get an idea of the student's background using a computer. It also queried their at-home access to computing devices. The students answered questions like "What type of operating system do you use?", rating their experience with computer and web-based applications, and to briefly discuss their exposure to programming. The results of this survey can be found in Table 7 and Appendix 3.3, along with a discussion of the survey design and its implications. This survey characterized the class in an important way. Most of the students (91%) reported that they were either "comfortable" with the computer or "very comfortable" with the computer. "Comfortable" was defined for the students as " I can install/delete programs and learn new applications on my own". "Very comfortable" was defined as "I'm the one everyone calls whenever the computer needs to be fixed". Three students reported they were "not comfortable—I can only do things if I'm shown how, and don't know what to do to fix it". While 86% of the students had little to no experience with programming, the second most common response to the question asking about their experience programming a computer was to talk about a two-week QBASIC activity they did in the seventh grade. This survey helps characterize the setting and population we are studying. Ninth grade students with little to no computation background are being tasked to build predictive motion models using computation. These students demonstrate competence in using computation as well as deep understanding of the physics involved.

## 1.6    Research Questions

One of the goals of the NRC framework is to make K-12 science education more like what professional science practice. This work has narrowed this goal to focus on the NRC framework's highlighted scientific practices of modeling and computation. How effective was the implementation described in section 1.5 at teaching students to use computation? The large majority of the student population in this study did not have any prior experience coding. Does this computational naiveté have an effect on the students' ability to learn to code? As a group, how successful were students at writing code in a test setting? What did successful students do when they wrote code? What did unsuccessful students do when they wrote code? Are students who write code memorizing algorithms or do they have some deeper understanding?

## 2    ASSESSING STUDENT USE OF COMPUTATION IN A HIGH SCHOOL PHYSICS CLASS

A version of this chapter appeared in the American Institute of Physics Conference Proceedings 1513 titled *Understanding Student Computational Thinking with Computational Modeling* [53].

### 2.1    Introduction

We have worked with an in-service high school physics teacher for the past two years to develop a computational curriculum for a 9th-grade conceptual physics course. The high school instructor has used the Modeling Instruction physics curriculum for several years at a private co-educational K-12 day school in Georgia. He has also presented simulations of physical phenomenon that were written using the VPython programming environment. VPython allows students to create three-dimensional simulations easily and to accompany those simulations with graphs and motion diagrams that update in real-time [50]. To facilitate instruction in computation, we have developed a suite of computational assignments (using VPython) that complement and enhance Modeling Instruction's treatment of force and motion topics [27]. The philosophy and motivation for these assignments are described in [27] and in the first chapter of this thesis. The original version of these assignments are available on the classroom teacher's blog and in appendix 3.2 [54]. These lab assignments have gone through several major revisions and the ones used in the pilot high school class are considerably different in some respects (e.g., data collection is now done via video cameras or smart phone cameras exclusively and analyzed in Tracker [55]). The motivation for these changes will be discussed in the conclusion.

During the fall semester 2011, high school students were instructed to develop computational models of four Modeling Instruction force and motion models (constant velocity, constant acceleration, balanced forces, and unbalanced forces) to predict the motion of objects described by various mathematical models. We confined our computational extensions to models described by Newton's 2nd Law because computational modeling highlights the similarities of these four models [27]. Students would observe different types of motion in class (e.g. constant velocity, constant acceleration, etc.) and then attempt to describe this motion with computation. In all computational activities, students used Euler-Cromer numerical integration to determine the velocity and position after each time step [52]. Students were also instructed to use the net force divided by mass in their program rather than simply the acceleration (e.g., `baseball.v = baseball.v + Fnet/baseball.m *deltaT`) to update the velocity. This emphasized the force's relationship to the equations of motion [27].

It is important to note that students were not tasked with writing programs "from scratch." They were given scaffolded code that would have many of the computational components written for them like class declaration, importing modules, etc. They were then instructed to focus on the missing physics parts of the python code. This would include the creation of the Euler-Cromer step integration of Newton's 2nd law as well as initial conditions and any other physics necessary for the program to represent the observation. An example of the scaffolded program can be found in appendices 3.2 and 3.4.

Computational assignments followed in-class experiments and problem-solving sessions. Students participating in in-class experiments while exploring the constant velocity model obtained and graphed data from battery powered cars. Students then constructed a computational model of a constant velocity car. Students used these computational models to reproduce their experimental data and, later, to make predictions for a variety of physical situations to which the model applied. Students participating in problem-solving sessions worked collaboratively in small groups on problems in an instructor created packet. The instructor would circulate the room interacting with students. Later they would present problems to each other on whiteboards. During the problem presentations the students would play the "Mistake Game" [56]. Students would intentionally introduce mistakes into their work. Then the audience (i.e. other students and the instructor) asked the presenter questions to help the presenter discover their mistake.

We implemented computational instruction in two separate 9th-grade physics classrooms (36 students were in the course total). Each student had access to VPython on a laptop. Students also used the Georgia Tech-developed Python module PhysUtil [57]. PhysUtil was designed specifically to support the Modeling Instruction curriculum, and allows students to create graphs, motion diagrams, axes, and timers by writing only one or two lines of code. PhysUtil has since been augmented to utilize the CSV module offered in the Python standard library [16]. This allows easy comparison of experimental data and model prediction in python or to output model data.

Students' use of computation was evaluated with three separate assessments. Firstly, students attempted to develop a computational model of a physics problem using VPython in a proctored environment. Through this proctored assignment, we assessed whether students were capable of writing a VPython program without any aids. Success in this matter alone does not necessarily constitute success in modeling the physical system; students can write syntactically correct programs with incorrect physics. Analysis of students' code provided a cursory view of the types of challenges (whether syntactical or physical) the students faced when constructing a computational model. While it is important for students to write programs correctly, programming is not computational thinking [58]. To probe their reasoning, students were asked to complete a second assessment by answering an essay question designed to ascertain how they connected their computational model to the physics that the model described. Thirty two students completed this assessment. In particular, they were asked to describe how their computational model related to the physical model via the iterative loop. Analysis of the essay responses indicated that we needed to delve more deeply into student reasoning. Therefore, a subset of five students was selected to participate in a final think-aloud interview in which they described how to develop a computational model for a particular physical phenomenon. To provide a representative sample of students, we selected participants from a cross-section of different performance levels on the previous assignments.

## 2.2    Proctored assignment

For the proctored assignment, students attempted to develop a 2D computational model that determined the location and velocity of a thrown baseball after a specified amount of time. Students completed this model individually and without aid from their instructor. The proctored assignment was deployed on an online homework system that can randomize the values given in problem statements. Students were provided with a program scaffold that imported the necessary modules, created the baseball and ground objects, and defined the integration loop structure. See appendix 3.4 for an example of a program scaffold and a completed assignment. To complete the assignment successfully, students would assign the appropriate initial conditions and complete the integration loop by employing Euler-Cromer integration [52]. To facilitate students' successful completion of this assignment, students were given a "Code Checking Case" [59]. In the "Code Checking Case", students were provided with the correct final position and velocity of the ball after the given time had elapsed. Students could use this case to check if their program modeled the situation correctly. After completing the "Code Checking Case" students modeled a similar physical situation for the "Grading Case". In the "Grading Case", the initial conditions were altered including the integration time and the system was moved from the Earth to the surface of the moon to reduce gravity. Answers were not provided for the "Grading Case." Students input their final answers for the baseball's final location and velocity and uploaded their code to the homework system.

We sought to determine students' success rates and if their struggles were due to challenges with physics or with computational modeling. We started with codes that were developed to analyze student code of orbital motion in the introductory physics class at Georgia Tech [59]. These codes are split into five groups concerning different concepts: initial conditions, force calculation, updating Newton's 2nd law, and "other" concerning coding errors. New codes were suggested during the coding process. The entire code set can be found in Table 8.

Table 8: Codes validated in previous work used for the proctored assignment [59]. Codes marked with asterisks(*) were not applicable to the proctored assignment. GT students who were tasked with modeling the motion of planets became confused when trying to select the appropriate exponent of the interaction constant [59]. This issue does not arise when modeling projectile motion and thus IC5 was not needed. In orbital motion, an explicit separation vector is required due to the nature of motion (i.e., where is the separation vector and not necessarily zero), GT students attempted to raise this to a power (when in fact it is the magnitude of the difference that should be raised not the vector) [59]. In projectile motion the separation vector is always zero therefore there is never a need to interact with it (this is the excluded FC3). Due to the simplicity of the forces involved in the proctored assignment students didn't have any other force direction confusion (the only force direction would be the direction of gravity) thus FC5 was inapplicable. Using Euler-Cromer integration to represent Newton's 2nd law programmatically, students modeling orbital motion often had difficulties differentiating between when to use a scalar or a vector [59]. Since the acceleration is one dimensional in projectile motion SL3 was not needed. SL4 was in regards to momentum updates and the proctored assignment does not consider momentum in its calculations (students were told force is not the more general ) [59]. In projectile motion, the "massive particle" is the Earth. Generally, Newton's theory of gravitation will work for this situation. However this can be reduced to for projectile motion and this is done for the proctored assignment removing the need for O1.

| Code | Description |
| --- | --- |
| IC1 | Student used all the correct given values from the grading case |
| IC2 | Student used all the correct values from the test case |
| IC3 | Student used the correct integration time from either the grading case or test case. |
| IC4 | Student used mixed initial conditions |
| IC5* | Students confused the exponents on the units of the exponent of k (interaction constant). |
| FC1 | The force calculation was correct |
| FC2 | The force calculation was incorrect, but the calculation procedure was evident. |
| FC3* | The student attempted to raise the separation vector to a power |
| FC4 | The direction of the force was reversed |
| SL1 | Newton's second law was correct |
| SL2 | Newton's second law was incorrect but in a form that updates |
| SL3* | Newton's second law was incorrect and the student attempted to update it with a scalar force |
| SL4* | Student created a new variable for p_f (momentum) |
| O1* | Student attempted to update force, momentum, or position for the massive particle. |
| O2 | Student did not attempt the problem |
| O3 | Student did not print final results |
| O4 | Coding Error [60] |
| SUG1 | Student didn't change anything inside the while loop |
| SUG2 | Student created extra while loops |
| SUG3 | Student printed in while loop |
| SUG4 | Student did not indicate Newton's 2nd law causes a change in motion |

Several of the codes needed to be dropped because they didn't apply, specifically, some issues were not seen due to the high school students having an easier assessment problem. The Georgia Tech (GT) students modeled the motion of orbiting planets, the high school students modeled the motion of a baseball flying through the air. Two of these codes deal explicitly with vector issues that only arise with complex problems like orbital motion (FC3, SL3). GT students were required to model momentum which was not required for high school students, thus O1 was jettisoned. Finally, the GT students needed to define the exponent of the separation constant which is not necessary for projectile motion (code IC5 and FC3).

The process for applying these codes had multiple steps and was based on producing high corre-lation across two coders. Working with another graduate student, a small selection of proctored as-signments were first analyzed using the codes from previous Georgia Tech work [59]. Examples and def-initions of each code were developed from a subset of the high school student coding assignments. The-se codes were applied to the entire data set. Cross correlation between codes at this stage indicated a low agreement across all of the data set so the smaller data set selection was revisited. This process was repeated until a high agreement between coders was reached. This process also generated new codes labeled SUG# in Table 8. SUG1-4 all arose from how students wrote code within the while loop. SUG1-3 are computational idiosyncrasies that were interesting and merit further exploration. SUG4 is interesting because students were explicitly instructed to define accelerations in their code in terms of force. Thus SUG4 happens when students eschew this for simply writing the acceleration, in this case $g = 9.8 \, \dfrac{\mathrm{m}}{\mathrm{s}^2}$ or $g = \langle 0, 9.8, 0 \rangle \, \dfrac{\mathrm{m}}{\mathrm{s}^2}$.

To calculate the agreement between coders we tabulated the total number of agreements we had (that is, the total number of times that two coders agreed that a code applied). Then we counted the number of total codes. For example, if researcher A gave three unique codes to a student, and re-searcher B gave two unique codes to a student, and researcher B's codes were contained within re-searcher A's codes than they would disagree 1/3 of the time. Likewise, if researcher A gave two unique codes and researcher B gave three unique codes but they only agreed once, the student would receive four unique codes total and the researchers disagreed twice. The ratio of the total number of agreements to the total number of unique codes gives the fractional agreement (that is, the inter-rater reliability). In this case researchers agreed 83.8% of the time [61].

Our analysis of student python code suggests that high school students can engage in computational modeling in the context of physics and that these students are generally capable of using numerical computation to solve physics problems. Based on the output of their computational models, students were placed into one of three categories: "correct results and animation" (N = 10, 31%), "produced animation, but incorrect results" (N=8, 25%), and "produced no animation" (N=14, 44%). The group that produced an animation but had incorrect results could be broken down into two groups. They either produced some number of errors either writing the integration algorithm alone (25%) or writing the integration algorithm and assigning initial conditions (75%). The remaining 44% who produced no animation were split into two groups. They either had small syntactic errors (36%) or had numerous physics and computational errors (64%). The group that had small syntactic errors most often would have succeeded with minimal input from peers or instructors because these errors were minute (e.g., missing a colon at end of while loop declaration but the rest of the program is correct).

The success rate here is important to note for several factors. First, the students had only spent the fall semester learning to use computation. During this semester students only had two full activities that had computation integrated. They then had a month long holiday break where they did not participate in any classroom activities. Upon returning the students had two weeks of beginning of year examinations that did not include any material on computation. After this two weeks the students had a short refresher (<1 day) on the Fall semester's material and then participated in the proctored assignment. Thus students, had at least a 6 week gap between the education period (i.e., the fall semester where they were learning to use computation in the context of force and motion) and the testing period. Four of these weeks were not spent participating in educational instruction at all. Furthermore, in an informal survey given by the classroom teacher indicates most students feel comfortable with the computer but have no programming experience with the computer. Thus, high school students with little programming experience can spend a relatively short period of time learning computation and a sizeable fraction will be able to apply this knowledge in context at a later time. With more instruction and more integration, a more favorable portion of the class could complete this assessment successfully. But do these students who complete code successfully understand it's link to the physics they are modeling? Or have they simply memorized algorithms that can only be used in very specific contexts.

## 2.3    Essay questions

The code that students wrote for the proctored assignment demonstrated a variety of possible outcomes, but this assessment was unable to probe deeply how students constructed these computational models. To look more deeply we asked the students to write a short essay.  Students responded to this essay question after they completed the proctored assignment. The essay question investigated whether students' success was predicated on simply reproducing an algorithm, or whether successful students made deeper connections between the physics and the computational algorithm. That is, did these students engage in the practice of computational thinking while developing their computational model? The question given to the students probed the student understanding of how the computer interacts with Newton's 2nd law via the Euler-Cromer integration. Twenty-nine of 32 students completed the essay question. Students could run a working version of the program before answering the essay question.

The practice of computational thinking requires a logical problem solving approach that often involves thinking iteratively [59]. To further investigate how students developed their computational models, we asked students to describe the integration (while) loop mathematically, physically, and programmatically. Specifically, the students were asked, "Download and run the completed baseball.py program. What is the purpose of the loop? How can you describe (mathematically or physically) what the loop does in your program? Run the program again and explain what the loop is doing while your program is running." In order to provide a complete explanation, students needed to comment on the iterative procedure of the loop itself and its relationship to the integration of the equations of motion by the incremental stepping of Newton's Second Law.

Three researchers catalogued all views that "arose" from the student responses. These views formed codes that were tested against a small subset (<5) of essays. This process was repeated until an agreement was formed on which codes applied where within the subset of essays. These codes were then applied to all of the submitted essays. This process was repeated until there was a high agreement between coders (100%).

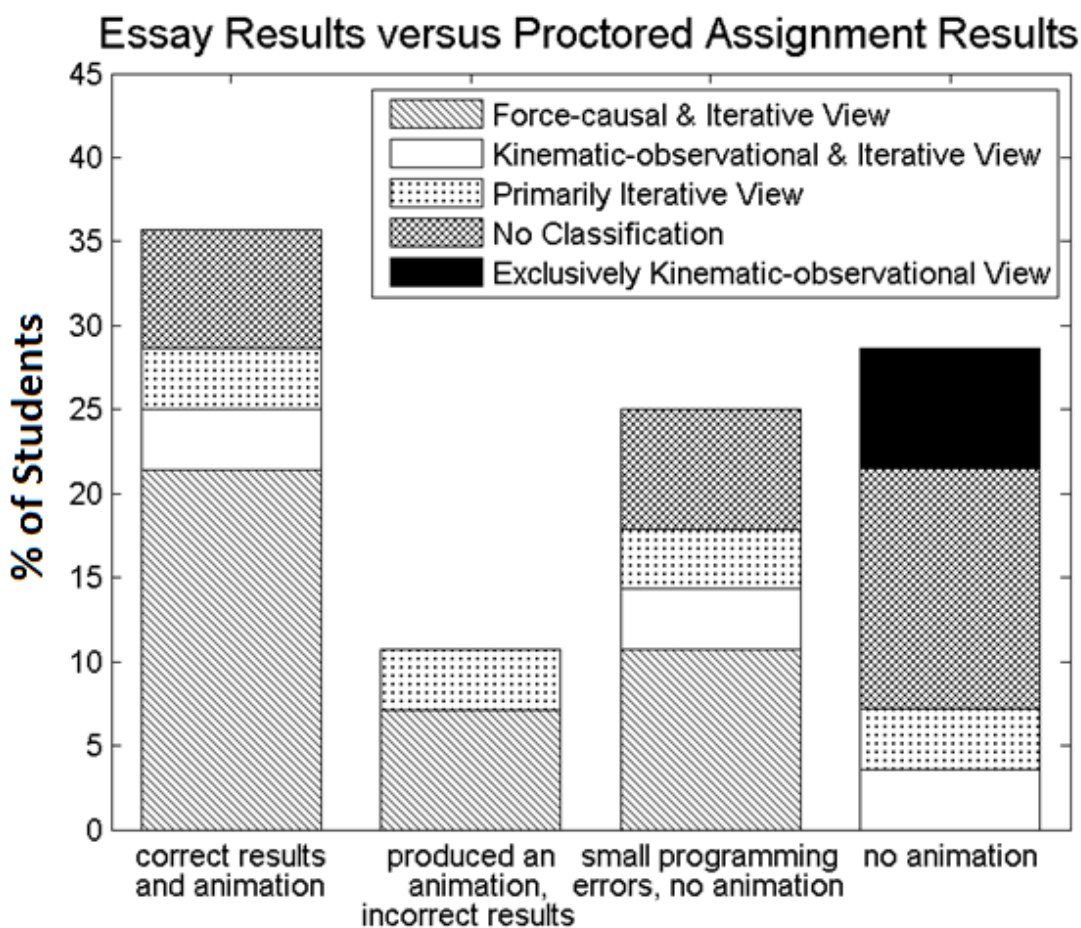

Figure 5: Students who displayed Force-casual and Iterative views were more likely to be successful on the proctored assignment (N=29). The bars represent a total proportion of the class (that is out of 100%). The percentages in this graph are **different than the proctored assignment because the number of students who responded to the essays is slightly less (proctored assignment N=32).**

The explanations presented by students in response to this question were captured by four distinct but not necessarily exclusive views. Some (38%) students presented a "force-causal" view of the loop structure. This view was characterized by a clear connection between force and motion. A student presenting a force-causal view would describe how the force of gravity causes a change in the motion of the ball; "The loop is constantly *changing the velocity of the ball while the Fnet [net force] stays constant. [The force] makes the ball fall faster* with every loop that runs". Another group (17%) of students presented a "kinematic-observational" view of the loop structure. These students indicated they had observed an acceleration (or some change in a kinematic quantity), but these students did not connect this observation back to the concept of a non-zero net force. One student with a kinematic-observational view noted, "The loop's purpose is to use the *acceleration of the ball to affect the ball's velocity and position.* The loop is run every .01 seconds (deltat). It re-updates the velocity and position of the ball at that interval." Almost two-thirds (65%) of students described the integration loop as a local, iterative process governed by instantaneous influences. This iterative-local view was characterized by a discussion of incremental steps through the loop and statements such as "in this program, the [integration*] loop is what the computer runs through* to [compute] a *new position, velocity, and all other forces for every [time it executes].*" All the students who exhibited a force-causal view and nearly all students who presented a kinematic-observational view of motion also exhibited an iterative-local view of motion. Slightly more than a quarter (28%) of all respondents fell into no category. This group of students most often wrote very short, incomplete responses that were too difficult to classify. More examples of student essay coding can be found in Table 9.

Table 9:Student views with examples. Underlined portions indicate sections coded as applying to the associated view. Examples might include multiple views but this in not included in these examples.

| Student View | Example |
|---|---|
| Iterative procedures | The program goes through all the steps and prints/runs them while the "while" statement is still true. While the program is running the loop is going through the steps very fast. |
| Force-Causal (i.e., mentions force and motion quantities together) | In this program the loop is what the computer runs through so put in a new position, velocity and all other forces for every .01 seconds. It is similar to a movie in that in a movie there are a bunch of pictures that scroll so quickly it looks like it is moving. When it is running it moves the ball and updates the time. |
| Kinematic-observational | The loop shows the changes in every vector as the time changes. I didn't really finish the vpython module so I don't completely understand everything yet. It also labels where the ball is at each second. |
| No category | No example |

We compared the views that students presented on the essay question to their performance on proctored coding assignment. Students with each view were binned into the broad proctored assignment categories (i.e., "correct results and animation", "produced animation, but incorrect results", and "produced no animation"). Students who presented both an iterative-local and force-causal view were most likely to produce a correct program. Students whose essay were short and incomplete were most likely to write programs that produced no animations. Figure 5 summarizes our findings.

## 2.4   High School Student Interviews

Students' essay responses elucidated that the concepts of force, motion, and iterative processes should be connected to facilitate computational thinking. However, investigating how students make these connections requires observing and questioning students while they engage in the practice of computational thinking. Several weeks after students completed the essay question, we interviewed five of them while they filled in the missing pieces of a scaffolded computational modeling program on paper. During the interview, students also answered questions about how they define a force and how forces, motion, and the integration loop are related. Students were asked to speak out loud while completing the scaffolded code and answering questions; their responses were videotaped. Only students whose proctored assignment code produced animations (i.e., "correct results and animation" and "produced animation, but incorrect results") were invited to the study. Six students were chosen to participate; five completed the interview. Of the students who completed the interview, 3 presented force-causal and iterative-local views on the essay question. One student had previously presented both a kinematic-observational and an iterative-local view, but expressed a force-causal and an iterative-local view in the interview. The last student presented a primarily iterative-local view on the essay question and in the interview.

For students who developed a correct computational model, the interviews further highlighted the links they made between force, motion, and iterative processes. A student who wrote a correct program described her code with a force-causal and an iterative-local view, "To *predict the velocity* you would have to *do baseball.v = initial velocity of the baseball plus gravity times time*. That would give me the new velocity *after [the execution of] every single loop*. And then you need *to update the position based on the loop*." This student presents the basic concepts behind Newton's 2nd law but also describes how the numerical integration loop updates the velocity of the ball in each execution. By contrast, a student who constructed a model that produced incorrect animation demonstrated an incorrect conception of force and motion, "*force generally [is] acquired through motion*. There's *always force acting* on an object." When questioned about how the loop models the physics of the system, the student presented solely an iterative-local view, *"[the loop] has formulas that it solves for*, like, update position equals [`baseball.pos + baseball.v*deltat`]." While this student was able to generate a computational model for the proctored assignment that ran without (syntactic) errors, she did not use the correct physics to do so.

# 3    CONCLUSIONS

## 3.1    Discussion

The results of this preliminary study are promising. 9th grade students with little to no pro-gramming experience were able to complete physics computational models to some (~31% totally cor-rect) degree of proficiency. This raises the question, "What defines success?" In this implementation, there were many decisions made by the experimenters that impacted the overall success of the stu-dents. Students spent very little time in class actually programming. There were only two computational assignments given to the entire class. There was also no link between experiment and computational model building. These two choices are probable causes for the low "total success" number. However, given the time between the educational period and the assessment period (6-8 weeks) and the success rate (31% wrote completely correct programs), the results seem more positive than they would at first glance. The 31% success rate indicates that students in high school can learn a computational tool like python in the context of physics (with effort). This is reinforced by the fact that the majority of students (86%) had little to no background in programming. Thus, it behooves future researchers to carry out work that extends computation throughout the curriculum.

It is important to note the background of the students in this study. The high school in this study is a private day school in Georgia. While students were not directly asked for their parents income lev-els, some assumptions can be made about the socio-economic status of the students based on data col-lected with the technology survey. Twenty percent of the students had iPads (in 2011, the iPad had been out for less than two years and 8.9% of the population possessed one), 56% of the students had their own smartphone, 78% had their own laptop (60% of which were MacBooks), and all students had high-speed internet access at home [62]. While not a conclusive measure of socio-economic status, these data do suggest that students at the high school are affluent. Thus, their classroom performance could tend towards success more often than that of a student with lower socioeconomic status (SES) [63]. Fur-ther work must be done to make these curriculum transformations accessible to all students, regardless of socio-economic status.

Another variable that could affect student success is the classroom teacher's level of competence at computational modeling. Jackson reports that students are more successful with teachers that have been labeled "master modelers" in Modeling Instruction classrooms [9]. This reasoning could extend to teaching computational modeling as well. The classroom teacher has an MS in Physics with several years of experience in Python programming. He also had the support of a research group of PhD students, physics professors, and researchers whose focus was on teaching computational physics. Can less experienced teachers still see success in their classrooms with some training? What level of support will these teachers need? It stands to reason that initially these teachers will need previously generated activities and a framework to offer these activities. Community and departmental support will also be a must for these teachers.

One of the stated goals of teaching physics with computation is for students to understand that motion can be described with Newton's 2nd law. In Python code, initial conditions are first explicitly declared, then calculations are performed. If a student were to add another force to the model, this would be an explicit action on the student's part. Our analysis of the essays begins to show that students are thinking this way. Students who had a good understanding of how the computer interacted with the physics were better at producing successful code (see Figure 5). To further explore this connection, students should be challenged with a problem that is beyond their educational experience. For instance, one might give a student a problem to model with a mix of forces, like a ball rolling down a hill, falling into the ocean, then eventually settling on the ocean floor. If a student could a) describe the forces at each step, and b) describe the code that would model this situation, this could demonstrate change in a study which compares students in a traditional physics class to this group.

Students also need to see computation as a tool for solving physics. This question was not assessed in this research. Students' reactions to using computation is important because we want them to not only learn to use the tool, we want them to understand why it is important to learn the tool at all. Science class should be about practicing what scientists do. In conversations with classroom teachers at the pilot high school, students have described a disconnect between computation and physics [64]. This was an unfortunate outcome that may be due to the way computation was implemented in the MI curriculum.

## 3.2    Looking Back

Ninth grade students were exposed to the scientific practices of modeling and computation. Making science education more like science practice produces students better equipped to solve modern day problems [6, 25, 38]. Some students in this study demonstrated that in learning the scientific practices of modeling and computation, they extended their sense making. Students who competently produced programs were most likely to describe these models as a representation of a causal relationship between force and motion change. This corroborates previous research in the use of computation in the physics classroom [65-67]. High school students writing code also demonstrated problem-solving difficulties similar to the difficulties exhibited by undergraduate students [59]. Thus, academic maturity may not play a role in learning computation, at least in students who are adolescents or older. Confirming and expanding the results of this thesis will require more test classrooms and teacher support.

## 3.3    Looking Forward

Implementing course reform is never easy. Instructors report motivations for implementing research based instructional strategies ranging from "it was easy" to "motivated by evidence" [68]. Many instructors do not possess the skills to do computation themselves and need instruction to support implementing computation in their courses [11]. Integrating computation into a curriculum also requires the jettisoning of other knowledge or skills that could be covered [65]. Anecdotally, students buck at curriculum change, which can cause instructors to hesitate. This has been noted by teachers at the pilot high school [64]. Discussions with these teachers indicates this may be due to a poor link between computational modeling and predicting experimental observations.

In this implementation, the computational component was kept separate because it was only introduced to the students during the model deployment stage. The students would do an experiment, discuss it in class, and expand it with additional assignments. They then would work on a computational assignment that was related to the initial experiment. The classroom instructor would have a discussion about the predictive power of computation, but this would never be strongly linked back to the experiment. Future implementations should have a stronger link between the experiment/observation and the computational prediction. This would involve actually predicting experimental values with the computational model, then testing the accuracy of these models. This process would more closely resemble real scientific practice, and is currently being tested at the college level and with an online course [69].

Student sense-making while modeling with computation should be further developed. In this study, many successful students demonstrated that they did more than memorize an algorithm in their solution-making. How do these students organize this knowledge cognitively? How do they access it when they solve problems? If we can characterize this process, can we operationalize it and teach it to other students? Working with both high school students and college students, instructors note students often resist "tweaking" models. Students often stop at descriptions that are not physical. What factors in students knowledge's and skill sets lead them to wrong conclusions? How can we strengthen students meta-cognitive abilities that allow them to "tweak" more effectively?

**APPENDICES**

### 3.1    Reasons for Choosing Python as a Learning Language

**Table 10 : This table has been reproduced from [70].**

| Feature | Reason |
|---------|--------|
| **Small and clean syntax** | Compared to languages such as Java or C++, Python has a more intuitive syntax. Example Hello World program: |

Python:
```
Print "Hello World!"
```

Java:
```
class Hi {
public static void main (String
args[]) {
System.out.println("Hi!");
}
}
% javac Hi.java
% java Hi
```

| Feature | Reason |
|---------|--------|
| **Dynamic typing** | Python is dynamically typed, which further reduces the notation. |
| **Expressive semantics** | Python's basic types are powerful: for example, lists can be introduced at the same time as other built-in types. |
| **Immediate feedback** | The interpreter enables fast and interactive demonstration of programming concepts, and gives immediate and understandable feedback on potential errors. |
| **Enforced structural design** | Python enforces an indented and structured way of writing programs, and the code resembles pseudo code. |
| **Relevant open-source software** | Python is free and widely used. It comes with a text editor (IDLE), and a large amount of tutorials, books, course material, exercises, assignments and extensive documentation is available on the web. |

### 3.2    Computational Activities Designed and Used by John Burk

These activities can be found online at http://quantumprogress.wordpress.com/computational-

modeling/. While the activities report (at the bottom) that they were part of the 2006 Modeling Work-shop Project, they were not developed for this project and the only relation to the project is our work augmenting Modeling Instruction with computation.

### 3.2.1    Introduction to Computational Modeling

# Introduction to Computational Modeling

One of the most important tools you will learn this year is to build simulations of physical situations (like a car moving at constant velocity) that run on a computer using VPython, a very powerful programming language. It will take some time to fully myelinate your neurons to master this, but with curiosity and patience, you will be amazed by what you can do.

The first thing you need to do is see the big picture of why we would want to go through the trouble of learning to program in a physics class, and what computational modeling allows us to do.

I've created this video on the right give you an introduction to this idea.

1. Start by watching "The big idea of computational modeling" on the right

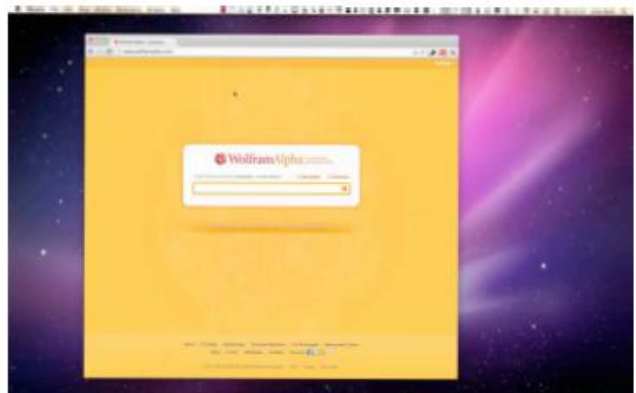

**Install VPython by following these directions:**
1. Go to www.vpython.org and follow the installation instructions for your particular platform.
2. Follow instructions for installing **VPython 2.7**
3. Download the program you will be working with by clicking on this link: bit.ly/unit1vp

**The big idea of computational modeling**

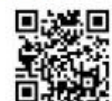
http://bit.ly/CVPMvid1

**Your assignment:**
4. Watch this video, "Getting started with VPython" on the right.

5. Open VIDLE, and the chose File > Open and navigate to the file **1-dMotionSimulation.py** *You can find an annotated version of the project on page 5 of this packet.*

6. **Modify the code that you downloaded in step 3 above so that the simulation exactly matches the motion of your buggy. Remember that the measurements in your program are in meters.**

7. Compete the assignment **Intro to Computational Modeling** on webassign.

- Additional Challenges:
    - Make the object start from the other end of the field and move in the opposite direction.
    - Make the object stop after 10 s.
    - Make the object turn around after 10 s.
    - Can you add a second cart to the program?

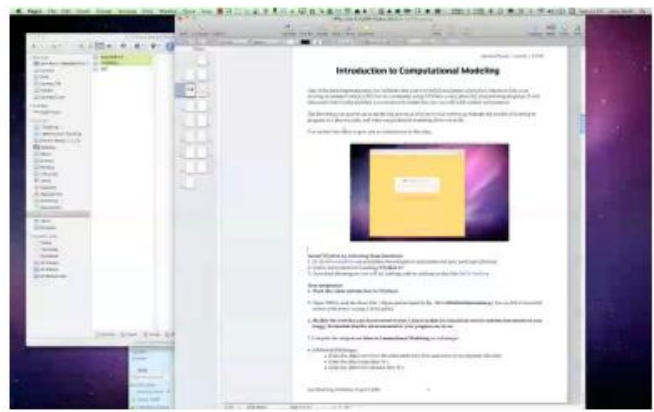

**Getting started with VPython**

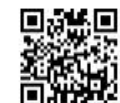
http://bit.ly/CVPMvid2

*from* Modeling Workshop Project © 2006          – 4

Honors Physics / Unit 01 / CVPM

```
### INITIALIZE VPYTHON
# ---------------------------------------------------------------

from __future__ import division
from visual import *
from physutil import *
from visual.graph import *
```

Set up the theater

```
### SETUP ELEMENTS FOR GRAPHING, SIMULATION, VISUALIZATION, TIMING
# ---------------------------------------------------------------

# Set window title
scene.title = "My Buggy Model"

# Make scene background black
scene.background=color.black

# Define scene objects
track = box(pos =vector(0,-.1,0),size=(3,.1,1),color = color.green) #units are m
car = box(size=(.3,.1,.2), color = color.blue)

# Define axis (with a specified length) that marks the track with a specified number of tick marks
axis = PhysAxis(track, 16, length=3) #units are m

# Set up graph
positiongraph = PhysGraph()

# Set up trail to mark the car's trajectory
trail = curve(color = color.yellow, radius = .01) #units in m

# Set timer in top right of screen
timerDisplay = PhysTimer(1,1)
```

Assemble the actors in the play

```
### SETUP PARAMETERS AND INITIAL CONDITIONS
# ---------------------------------------------------------------

# Define parameters
car.m = 1. #mass of car in kg
car.pos = vector(0,0,0) #initial position of the car in (x,y,z) form, units are m
car.v = vector(-.5,0,0) #initial velocity of car in      units are m/s

# Define time parameters
t=0 #starting time
deltat = 0.001  #time step units are s
```

Set up actors on stage at beginning

Physics is here

```
### CALCULATION LOOP; perform physics updates and drawing
# ---------------------------------------------------------------

while  car.pos.x > -1.50 and car.pos.x < 1.50 :  #while the ball's x-position is between -1.5 and 1.5

    # Required to make animation visible / refresh smoothly (keeps program from running faster than 1000 frames/s)
    rate(1000)

    # Compute Net Force
    Fnet = vector(0,0,0)

    # Newton's 2nd Law
    car.v = car.v + Fnet/car.m * deltat

    # Position update
    car.pos = car.pos + car.v*deltat

    # Update timer, graph, and trail
    timerDisplay.update(t)
    positiongraph.plot(t,car.pos.x)  #this plots one point in the graph in (x,y) form
    trail.append(pos = car.pos)

    # Time update
    t=t+deltat
```

Physics is here

Run the play

```
### OUTPUT
# ---------------------------------------------------------------

# Print the final time and the car's final position
print t
print car.pos
```

Review the results

### 3.2.2   WebAssign Constant Velocity Assignment

WebAssign is an online homework system.

---

**Vpython CVPM Assignment 1 (1847873)**

| Question | 1 | 2 | 3 |

**Description**
The purpose of this assignment is to check your understanding of computational modeling and vpython.

---

**1.    Question Details**                                                    VP CVPM 1.2 [1774929]

If you want to make the object start at the right side of the field and move to the left, in which section of the program do you need to make a change?

- ○ Setup Elements
- ○ Initialization
- ○ Initial Conditions
- ○ Calculation loop
- ○ Post Loop/Output

---

**2.    Question Details**                                                    uI vpyhton 1 [1863255]

Input the velocity and starting position for your buggy for your program:

velocity  [________________]  m/s
starting position  [____________________________]  m

Enter the line of code you must change in order to make the velocity of the car in the program match the velocity of your buggy.  [________________________]  Enter the line of code you must change in order to make the starting position of the car in the program match the starting position of your buggy.
[________________________]

---

**3.    Question Details**                                                    CVPM 3 (prediction) [1863259]

Suppose that your buggy travels at a speed of 0.25 m/s, and you wish to use your program to determine how much time it would take for the buggy to travel from Westminster to Lovett, a distance of 2.7x10^3 m (or 2.7E3 in python notation).

Describe how you could use your program to determine the time it takes for the buggy to make this trip, without using a calculator or doing any math in your head. In your description describe what lines of code you would need to modify and why.

[________________________________________________]

Now try to modify this code to test your method, and find the distance. Note: You can significantly speed up your program by commenting out the line inside your loop that reads:

rate(1000)

Simply put a # symbol in front of this line, and it will let the program run as fast as possible.

You can also speed the program up by making the time step longer in the initial conditions section. (try 1 instead of 0.01)

deltat 1

Still, your program will take a significant amount of time to run, so it is best to start this program and go and work on something else while you wait.

[ ] seconds

**Assignment Details**

Computational Modeling Activity 2: Balanced Forces

# Computational Modeling Activity 2: Balanced Forces

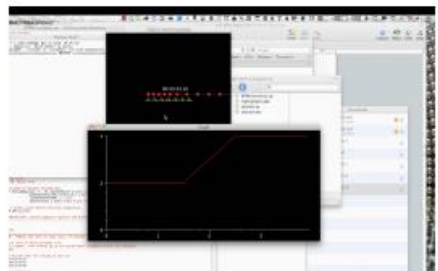

1. Begin by downloading the assignment from http://bit.ly/unit2vpython. If you are running this on PC, you will need to extract the files from the zip archive (instructions: http://d.pr/ewPu)

2. Run VIDLE and open the program BFPMsimulation.py.

3. Watch the video linked on the right for a short introduction to this assignment. This program creates an object with an initial velocity to the right, and then allows you to change the forces acting on it at various points in the program (Link to video http://bit.ly/BFPMvid)

4. Use this simulation to explore how the motion of an object changes as you change the forces acting on the object. *Remember that you must click the mouse to start the simulation.*

5. Conduct an investigation using this project to determine the answer to this question: "How can you tell, based only on the motion of an object, whether the net force acting on the object is zero?"

6. Use your program to answer this question: "If an object is moving to the right, must it also be experiencing a force to the right?"

7. Now suppose that the following two forces are acting on the object (paste this code in at line 110, just after the end of loop 1, replacing the two lines that are already there):

```
F1=vector(0,2,0) #force defined in (Fx, Fy, Fz)
F2=vector(2,1,0)
```

Find the value for **F3** needed to balance forces and create a net force of 0.

8. Answer each of these questions in the BFPMVPython on Webassign.

**Extensions/Challenges**

1. You've probably noticed that this code is rather long and messy. There are lots of ways to clean it up, but one easy method is to take blocks of the code (for instance, everything inside the setup block, and put it into a python file, named setup.py). Then add the line of code

```
from setup import *
```

use this method to clean up the setup section of the code.

2. You can also also considerably shorten blocks of code by using functions, which allow you to execute multiple statements with a single call (many of the things you are using now are functions). Try adding this function to the setup block of your code

```
def featureUpdate(posgraph, velgraph, timer, Map, t, car):
        posgraph.plot(t,car.pos.x )
        velgraph.plot(t,car.velocity.x)
        timer.update(t)
        Map.update(t)
```

then replace the statements below the comment #update graph and simulation features with the line:

```
featureUpdate(postiongraph, velocitygraph, timerDisplay, motionMap, t, car)
```

How else can you use functions to clean up and improve this program?

Honors Physics / Unit 02 / BFPM

```
### INITIALIZE VPYTHON
# ----------------------------------------------------------------

from __future__ import division
from visual import *
from physutil import *
from visual.graph import *

### SETUP ELEMENTS FOR GRAPHING, SIMULATION, VISUALIZATION, TIMING
# ----------------------------------------------------------------

# Set up simulation window
scene.title = "FORCES INVESTIGATION"

# Define scene objects
track = box(pos =vector(0,-1,0),size=(6,.1,1),color = color.black) #units are m track is black so that it is invisible
car = box(size=vector(.3,.1,.2), color = color.blue, make_trail=True) #the make trail command causes the object to display trail

#create arrows to represent individual forces
F1arrow=arrow(axis=vector(0,0,0),color=color.yellow)
F2arrow=arrow(axis=vector(0,0,0),color=color.yellow)
F3arrow=arrow(axis=vector(0,0,0),color=color.yellow)
Fnetarrow=arrow(pos=vector(0,1,0),axis=vector(0,0,0),color=color.red)

# Define axis (with a specified length) that marks the track with a specified number of tick marks
axis = PhysAxis(track, 7,axisColor=color.white)

# Set timer in top right of screen
timerDisplay = PhysTimer(1,1)

# Set up graphs in separate windows
positiongraph = PhysGraph()
velocitygraph = PhysGraph()

### SETUP PARAMETERS AND INITIAL CONDITIONS
# ----------------------------------------------------------------

# Define parameters
car.m=0.5 #mass of car
car.pos = vector(-3,0,2,0)  #initial position of the car in( x,y,z) form, units are m
car.velocity = vector(2,0,0) #initial velocity of car in (vx,vy,vz) form, units are m/s

# Define time parameters
t=0 #starting time
deltat = 0.01  #time step

#Set up MotionMap to display breadcrumbs
motionMap = MotionMap(car, 6, 15, markerType="breadcrumbs",   #drop 6 breadcrumbs between t=0 and t = 15
                      labelMarkerOffset=vector(0,1,0,0), #put lables below the marker
                      labelMarkerOrder = False,
                      markerColor = color.red) # put times above the marker

# Wait for a mouse click before starting simulation
scene.mouse.getclick()

### CALCULATION LOOP; perform physics updates and drawing
# ----------------------------------------------------------------

##Begin LOOP1
#
while t<1.5:  #while the time is less than 1.5 seconds

    # Control rate at which program runs
    #(larger number, runs faster up to the point where hardware limits are reached)
    rate(100)

    #define forces that are acting on the car
    F1=vector(0,0,0)
    F2=vector(0,0,0)
    F3=vector(0,0,0)

    # Compute Net Force
    Fnet = F1+F2+F3

    #define the lengths of the force vectors
    F1arrow.axis=F1 #the arrow representing F1 has the length of the F1 force
    F2arrow.axis=F2
    F3arrow.axis=F3
    Fnetarrow.axis=Fnet

    # Apply Newton's 2nd Law to predict new velocity
    car.velocity = car.velocity + (Fnet/car.m)*deltat

    # With new velocity, predict new position
    car.pos = car.pos + (car.velocity)*deltat)

    #update the location of the force vectors to match the position of the car
    F1arrow.pos=car.pos
    F2arrow.pos=car.pos
    F3arrow.pos=car.pos
    Fnetarrow.pos=Fnet+vector(0,1,0)

    # Update graph and simulation features (timer, motion map, etc.)
    positiongraph.plot(t,car.pos.x ) #this plots one point in the graph in (t,x) form
    velocitygraph.plot(t,car.velocity.x)
    timerDisplay.update(t)
    motionMap.update(t)   #updates the motion map (bread crumbs)
```

**Physics Here**

**Physics Here**

– 11 –

*from* Modeling Workshop Project © 2006

```
    # Time update
    t = t + deltat

#END Loop 1
# --------------------------------------------------

#change the values of the forces
F1=vector(1,0,0) #force defind in (Fx, Fy, Fz)
F2=vector(1,0,0)
F3=vector(0,0,0)

#Begin Loop 2
# --------------------------------------------------

while t<2.5 :   #while the time is less than 2.5 seconds

    # Control rate at which program runs
    #(larger number, runs faster up to the point where hardware limits are reached)
    rate(100)

    # Compute Net Force
    Fnet = F1+F2+F3

    #define the lengths of the force vectors
    F1arrow.axis=F1 #the arrow representing F1 has the length of the F1 force
    F2arrow.axis=F2
    F3arrow.axis=F3
    Fnetarrow.axis=Fnet

    # Apply Newton's 2nd Law to predict new velocity
    car.velocity = car.velocity + (Fnet/car.m)*deltat

    # With new velocity, predict new position
    car.pos = car.pos + (car.velocity*deltat)

    #update the location of the force vectors to match the position of the car
    F1arrow.pos=car.pos
    F2arrow.pos=car.pos
    F3arrow.pos=car.pos
    Fnetarrow.pos=car.pos+vector(0,1,0)

    # Update graph and simulation features (timer, motion map, etc.)
    positiongraph.plot(t,car.pos.x )  #this plots one point in the graph in (t,x) form
    velocitygraph.plot(t,car.velocity.x)
    timerDisplay.update(t)
    motionMap.update(t)    #updates the motion map (bread crumbs)

    # Time update
    t = t + deltat

#END Loop 2
# --------------------------------------------------

#change the values of the forces
F1=vector(0,0,0) #force defind in (Fx, Fy, Fz)
F2=vector(0,0,0)
F3=vector(0,0,0)

#BEGIN Loop 3
# --------------------------------------------------
while t<4:   ##while the time is less than 4 seconds

    # Control rate at which program runs
    #(larger number, runs faster up to the point where hardware limits are reached)
    rate(100)

    # Compute Net Force (no actual computation since force is constant)
    Fnet = F1+F2+F3

    #define the lengths of the force vectors
    F1arrow.axis=F1 #the arrow representing F1 has the length of the F1 force
    F2arrow.axis=F2
    F3arrow.axis=F3
    Fnetarrow.axis=Fnet

    # Apply Newton's 2nd Law to predict new velocity
    car.velocity = car.velocity + (Fnet/car.m)*deltat

    # With new velocity, predict new position
    car.pos = car.pos + (car.velocity*deltat)

    #update the location of the force vectors to match the position of the car
    F1arrow.pos=car.pos
    F2arrow.pos=car.pos
    F3arrow.pos=car.pos
    Fnetarrow.pos=car.pos

    # Update graph and simulation features (timer, motion map, etc.)
    positiongraph.plot(t,car.pos.x )  #this plots one point in the graph in (t,x) form
    velocitygraph.plot(t,car.velocity.x)
    timerDisplay.update(t)
    motionMap.update(t)    #updates the motion map (bread crumbs)

    # Time update
    t = t + deltat
#END Loop 3
# --------------------------------------------------

### OUTPUT
# --------------------------------------------------------------------------------
print t
print car.pos
```

**Change Forces**

**Physics Here**

**Change Forces**

**Physics Here**

### 3.3    Student Technology Survey Results

Students were given an informal technology survey at the beginning of the school year. All students had high speed internet access at home. In addition to the school provided laptops 24 (out of 36) students had laptops they brought to school with them every day.

While this is an informal survey several comments can be made about the results due to the design. The question "how comfortable are you with the computer?" has a three point scale response where the vast majority of the students picked the middle response. Survey design research has show that on three point scales respondents often choose the middle response over lowest or highest response. Thus, the design of this question may have biased the students. The granularity produced by the question "Do you know anything about programming a computer?" is poor with a "Yes/No" response. The question is also vague. A more appropriate way of gaining this information may be to split the question into two parts. First ask the students to rate how comfortable they feel about their knowledge of writing computer code on a five point scale (1 is very uncomfortable, 5 is very comfortable). Then, ask the students to rate their experience with writing computer code on a five point scale (1 is no experience, 5 is expert level experience).

Only 20 out of 36 students responded "yes" to the question "Do you know anything about programming a computer?" (in this context, "programming a computer" means writing code in any computer language to produce a novel program or solution). In the follow up question asking students to explain their experiences only one reported that they had an advanced level of programming experience stating " I can code in Arduino [an embedded systems device] format very well and know bits and pieces of java and C++". The other group reported that they had very little prior exposure to programming or computation stating that the only experience was a brief period in 7th grade math using QBASIC (14). This group made statements like, " Very limited knowledge: QBasic in 7th glade". See Table 7 : Students were asked the question "Do you know anything about programming a computer?". If they responded yes they were asked to explain their experiences (N=36).

| Question | Response | |
|---|---|---|
| Do you have access to a computer high speed internet connection at home? | Yes | 36 |
| | No | 0 |
| Do you have a laptop computer that you can bring to school? | Yes, A Mac laptop | 15 |
| | Yes, a Windows laptop | 9 |
| | No, I don't have a laptop | 8 |
| | No, I have a laptop, but I can't bring it to school | 4 |
| Do you have a cell phone? | Yes | 35 |
| | No | 1 |
| Do you have a smartphone? | iPhone 4 | 7 |
| | iPhone 3G | 7 |
| | other iPhone | 1 |
| | Android Phone | 3 |
| | Blackberry | 1 |
| | Other smartphone | 1 |
| | No smartphone | 16 |
| Do you have an mp3 player? | iPod Touch | 20 |
| | iPod Nano | 2 |
| | Other iPod | 12 |
| | Other mp3 player | 1 |
| | No mp3 player | 1 |
| Do you have a tablet computing device? | iPad 1 | 6 |
| | iPad 2 | 1 |
| | Android tablet | 0 |
| | No | 29 |
| What type of operating system does your computer at home run? | Mac OS X | 16 |
| | Windows XP | 2 |
| | Windows Vista | 4 |
| | Windows 7 | 6 |
| How old is your computer? | <6 months | 7 |
| | 6 months - 1 year | 3 |
| | 1-2 years | 6 |
| | 2-3 years | 7 |
| | >3 years | 5 |
| How comfortable are you with a computer? | not comfortable—I can only do things if I'm shown how, and don't know what to do to fix it. | 3 |
| | comfortable—I can in- | 29 |

| | | | |
|---|---|---|---|
| | stall/delete programs and learn new applications on my own. | | |
| | very comfortable—I'm the one everyone calls whenever the computer needs to be fixed. | 4 | |
| **Do you know anything about programming a computer?** | Yes | 20 | |
| | No | 16 | |
| **Rate your experience with the following programs** | program | No experience to basic experience | advanced to expert experience |
| | Word Processor (MS Word, Pages) | 7 | 29 |
| | Spreadsheet (Excel) | 15 | 21 |
| | Photo/Video software | 25 | 11 |
| | Google Docs | 27 | 9 |
| **Please rate your familiarity with the following web technologies** | program | No experience to basic experience | advanced to expert experience |
| | Blogging | 28 | 8 |
| | Twitter | 31 | 5 |
| | Facebook | 18 | 18 |
| | Search | 8 | 28 |
| | Web Site Design | 32 | 4 |

## 3.4   Proctored Assignment

This appendix includes the scaffolded code that was given to the students during the proctored assignment assessment as well as an example of what a completed computational model would look like.

### 3.4.1   *Code That was given to students at the beginning of the assessment*

```python
from __future__ import division
from visual import *
from physutil import *
from visual.graph import *

###SETUP Elements
```

```
#-------------------------------------------------------------
scene.title = 'Hit a home run!'
field = box(pos=vector(0,0,0),size=(122,10,50),color = color.green,opacity =
0.3)
baseball = sphere(radius=.5,color=color.red)

###SETUP PARAMETERS & INITIAL CONDITIONS
#-------------------------------------------------------------
baseball.pos = (-61,1,0)
baseball.m=.145 #mass in kg
baseball.v=vector(25,25,0)
ballparkwall = 122 #distance to outfield wall in m
gravity=vector(0,-9.8,0)

t=0
deltat=.01
tf = 10

###MOTION MAP
#-------------------------------------------------------------
motionMap = MotionMap(baseball, tf, 10, markerType="breadcrumbs",
  labelMarkerOffset=vector(0,.3,0), dropTime=True)

###CALCULATION LOOP; perform physics updates and drawing
#-------------------------------------------------------------
while baseball.pos.x < ballparkwall and (baseball.pos.y)>field.pos.y: #while
the ball is inside the park

  rate(100)

  #Calculate forces

  #update velocity

  #update position
  baseball.pos = baseball.pos + baseball.v*deltat

  #update motionMap
  motionMap.update(t)

  #update time
  t=t+deltat

###OUTPUT
#-------------------------------------------------------------
print t
```

### 3.4.2   An example of successful code

```python
from __future__ import division
from visual import *
from physutil import *
from visual.graph import *

###SETUP Elements
#-----------------------------------------------------------
scene.title = 'Hit a home run!'
field = box(pos=vector(0,0,0),size=(122,10,50),color = color.green,opacity =
0.3)
baseball = sphere(radius=.5,color=color.red)

###SETUP PARAMETERS & INITIAL CONDITIONS
#-----------------------------------------------------------
baseball.pos = (-61,1,0)
baseball.m = .145 #mass in kg
baseball.v = vector(25,25,0)
ballparkwall = 122 #distance to outfield wall in m
gravity = vector(0,-9.8,0)

t=0
deltat=.01
tf = 10

###MOTION MAP
#-----------------------------------------------------------
motionMap = MotionMap(baseball, tf, 10, markerType="breadcrumbs",
  labelMarkerOffset=vector(0,.3,0), dropTime=True)

###CALCULATION LOOP; perform physics updates and drawing
#-----------------------------------------------------------
while baseball.pos.x < ballparkwall and (baseball.pos.y)>field.pos.y: #while
the ball is inside the park

  rate(100)

  #Calculate forces
  netforce = gravity*baseball.m #+ any other forces

  #update velocity
  baseball.v = baseball.v + netforce/baseball.m*deltat

  #update position
  baseball.pos = baseball.pos + baseball.v*deltat

  #update motionMap
  motionMap.update(t)

  #update time
  t=t+deltat
```

```
###OUTPUT
#------------------------------------------------------------
print t
print baseball.v
```